# Magnetic field dependence of critical currents of cross-type Josephson junctions with inhomogeneous critical current density under oblique magnetic fields

Soma Haraoka[1], Edmund Soji Otabe [2,3], and Yasunori Mawatari [4]*

[1] *Department of Computer Science and Systems Engineering, Kyushu Institute of Technology, Fukuoka 820-8502, Japan.*

[2] *Department of Physics and Information Technology, Kyushu Institute of Technology, Fukuoka 820-8502, Japan.*

[3] *Research Center for Neuromorphic AI Hardware, Kyushu Institute of Technology, Kitakyushu, Fukuoka 808-0135, Japan*

[4] *National Institute of Advanced Industrial Science and Technology (AIST), Ibaraki 305-8568, Japan*

*E-mail: y.mawatari@aist.go.jp

**Abstract**

Studies of the magnetic interference of sandwich-type Josephson junctions in which perpendicular or oblique magnetic fields are applied to the junction plane have received less attention than those where the applied magnetic fields are parallel. Recently, it has been theoretically demonstrated that a variety of magnetic interferences of the critical currents appear when oblique magnetic fields are applied to a cross-type junction with homogeneous critical current density. We theoretically investigated the effect of the inhomogeneous critical current density in the junction plane, and found that more complicated magnetic interferences appeared. We considered the distribution of the current density flowing through the junction plane to explore the cause of these complex magnetic interferences.

## 1. Introduction

The magnetic field dependence of the DC critical currents in sandwich-type Josephson junctions shows a Fraunhofer-type interference pattern when a magnetic field is applied parallel to the junction plane.[1–4] This magnetic interference is the basic principle for applications such as superconducting quantum interference devices.[2, 4–6] However, there are still only a few studies on magnetic interference in perpendicular or oblique magnetic fields. Rosenstein and Chen[7] as well as Hebard and Fulton[8] revealed that magnetic interference appears in overlap-type Josephson junctions under perpendicular magnetic fields. Miller et al.,[9] on the other hand, revealed that no magnetic interference appears in cross-type Josephson junctions under a perpendicular magnetic field. Monaco et al.[10,11] numerically and experimentally investigated the magnetic interference of various junction geometries under oblique magnetic fields. Furthermore, it was analytically demonstrated that anomalous magnetic interference patterns appear in cross-type junctions under oblique magnetic fields.[12]

The magnetic interference of Josephson junctions is also affected by the distribution of the critical current density $J_c$ in the junction area.[4,13] Analyses of magnetic interference of critical currents and inhomogeneity of $J_c$ have been used to investigate the edge states in topological insulators.[14–16] In contrast to the work in Ref. 12, which focuses on cross-type junctions with homogeneous critical current density $J_c$, we also consider the effect of $J_c$ inhomogeneity. We found that the magnetic interference patterns of cross-type junctions under oblique magnetic fields for inhomogeneous $J_c$ show more complex and diverse patterns than for homogeneous $J_c$.

This paper consists of five sections including this section. The next section describes the theoretical model we used to consider the magnetic interference of cross-type Josephson junctions. The third section presents the results of the magnetic interference in oblique magnetic fields. In the fourth section, we consider the difference in magnetic interference for homogeneous and inhomogeneous $J_c$ on the basis of the distribution of the current density flowing through the junction plane. The final section summarizes our results.

## 2. Theory

### 2.1 Cross-type junction

Figure 1 illustrates a cross-type Josephson junction consisting of two superconducting nanostrips and a barrier layer exposed to a magnetic field $(H_x, H_y, H_z)$. The top strip is along the $y$ axis, and the bottom strip is along the $x$ axis. The two superconducting strips have the same width $w$ and thickness $d_s$. The thickness of the barrier layer is $d_j$, and the junction area is a square measuring $w \times w$.

We assume that magnetic screening is weak in superconducting nanostrips (i.e., $d_s < \lambda_L$ and $w < \lambda_L^2/d_s$, where $\lambda_L$ is the London penetration depth) and in small junctions (i.e., $w < \lambda_J^2/\lambda_L$, where $\lambda_J$ is the Josephson length).[17] The effect of the self field due to the transport current is neglected.

2.2 Critical current

The current density flowing through the junction area $J_z(x, y)$ is given by

$$J_z(x, y) = J_c(x, y) \sin \theta(x, y), \tag{1}$$

where $J_c$ is the critical current density. The gauge-invariant phase difference $\theta$ is expressed as $\theta(x, y) = \theta_0 + \theta_1(x, y)$, where $\theta_0$ is a constant independent of the magnetic field and $\theta_1(x, y)$ for cross-type junctions is given by [9]

$$\theta_1(x, y) = \frac{2x}{w}\frac{\pi \Phi_y}{\phi_0} - \frac{2y}{w}\frac{\pi \Phi_x}{\phi_0} - \frac{2xy}{w^2}\frac{\pi \Phi_z}{\phi_0}, \tag{2}$$

where $\Phi_x = \mu_0 H_x w d_{\text{eff}}$, $\Phi_y = \mu_0 H_y w d_{\text{eff}}$, and $\Phi_z = \mu_0 H_z w^2$ are the magnetic fluxes in the junction, $\phi_0$ is the flux quantum, and $d_{\text{eff}}$ is the effective barrier thickness given by $d_{\text{eff}} = d_j + d_s$.[18] The net current $I_z$ flowing through the junction area is given by

$$\begin{aligned} I_z &= \int_{-w/2}^{w/2} dx \int_{-w/2}^{w/2} dy\, J_c(x, y) \sin \theta(x, y) \\ &= \int_{-w/2}^{w/2} dx \int_{-w/2}^{w/2} dy\, J_c(x, y) \sin[\theta_0 + \theta_1(x, y)] \\ &= \operatorname{Im}[F \exp(i\theta_0)], \end{aligned} \tag{3}$$

where $F$ is given by

$$F = \int_{-w/2}^{w/2} dx \int_{-w/2}^{w/2} dy\, J_c(x, y) \exp[i\theta_1(x, y)]. \tag{4}$$

The complex-valued function in Eq. (4) can be written as $F = |F|\exp(ig)$ with $g = \arg(F)$, and Eq. (3) yields

$$I_z = \operatorname{Im}\{|F|\exp[i(\theta_0 + g)]\} = |F|\sin(\theta_0 + g). \tag{5}$$

Maximizing $I_z$ with respect to $\theta_0$ yields $I_z = |F|$ when $\theta_0 = \pi/2 - g = \pi/2 - \arg(F)$, and the critical current $I_c = |F|$ as a function of $\Phi_x$, $\Phi_y$, and $\Phi_z$ is thus given by

$$I_c(\Phi_x, \Phi_y, \Phi_z) = \left| \int_{-w/2}^{w/2} dx \int_{-w/2}^{w/2} dy\, J_c(x,y) \exp[i\theta_1(x,y)] \right|. \tag{6}$$

The critical current at zero magnetic field $I_{c0} = I_c(0,0,0)$ is given by

$$I_{c0} = \left| \int_{-w/2}^{w/2} dx \int_{-w/2}^{w/2} dy\, J_c(x,y) \right|. \tag{7}$$

For cross-type junctions with homogeneous $J_c$, analytical equations for Eq. (6) are available.[12] The current density in Eq. (1) when $I_z = I_c$ [i.e., $\theta_0 = \pi/2 - \arg(F)$] is

$$J_z(x,y) = J_c(x,y) \cos[\theta_1(x,y) - \arg(F)]. \tag{8}$$

## 3. Results

### 3.1 Inhomogeneous critical current density

As a model for inhomogeneous $J_c(x,y)$, we consider the case where $J_c$ is piecewise constant. That is, $J_c = J_{c0}$ (where $J_{c0}$ is constant) in the outer region of $x_0 < |x| < w/2$ or $y_0 < |y| < w/2$, whereas $J_c = 0$ in the inner region of $|x| < x_0$ and $|y| < y_0$ in the junction area. The magnetic interference is investigated for four cases of the parameters $(x_0, y_0)$ shown in Fig. 2. For comparison, we also demonstrate the magnetic interference for the case of homogeneous $J_c$ investigated in Ref. 12 [i.e., for $(x_0, y_0) = (0,0)$ shown in Fig. 2(a)].

### 3.2 Critical currents under 2D oblique magnetic fields

Figure 3 depicts the critical currents of Eq. (6) as functions of $\Phi_x$ and $\Phi_z$ for $\Phi_y = 0$. The horizontal axis represents $\Phi_x/\phi_0$, which is the normalized parallel magnetic field. The vertical axis represents $\Phi_z/\phi_0$, which is the normalized perpendicular magnetic field. The color shade corresponds to the magnitude of $I_c$, where larger $I_c$ is indicated by lighter colors.

Figure 4 depicts the $I_c/I_{c0}$ as a function of $\Phi/\phi_0$ where $\Phi = \Phi_x$ or $\Phi = \Phi_z$. The blue line represents $I_c/I_{c0}$ vs $\Phi_x/\phi_0$ for $\Phi_y = \Phi_z = 0$, while the orange line represents $I_c/I_{c0}$ vs $\Phi_z/\phi_0$ for $\Phi_x = \Phi_y = 0$. Therefore, the blue line in Fig. 4 corresponds to $I_c$ on the line $\Phi_z/\phi_0 = 0$ in Fig. 3, and similarly, the orange line in Fig. 4 corresponds to $I_c$ on the line $\Phi_x/\phi_0 = 0$ in Fig. 3.

In the case of homogeneous $J_c$ [Fig. 3(a)], we see no stripe pattern of $I_c(\Phi_x, 0, \Phi_z)$ when $|\Phi_z| > 2|\Phi_x|$, indicating that magnetic interference does not occur.[12] On the other hand, in the case of inhomogeneous $J_c$ [Figs. 3(b)–(d)], magnetic interference occurs even

when $|\Phi_z| > 2|\Phi_x|$, and further unexpected magnetic interference is observed. Furthermore, in the case of $(x_0, y_0) = (0.4, 0.4)w$, a Fraunhofer-like interference appears under perpendicular magnetic fields as well [orange line in Fig. 4(b)], in contrast to the absence of the magnetic interference for homogeneous $J_c$ [orange line in Fig. 4(a)]. This phenomenon constitutes a critical difference between homogeneous $J_c$ and inhomogeneous $J_c$.

3.3 Critical currents under 3D oblique magnetic fields

Figure 5 depicts the critical currents as functions of $\Phi_x$, $\Phi_y$, and $\Phi_z$. The horizontal axis represents $\Phi_x/\phi_0$, and the vertical axis represents $\Phi_y/\phi_0$. The color legend for Fig. 5 is the same as that for Fig. 3. Figure 5 shows the cases where $\Phi_z/\phi_0$ is an integer ($\Phi_z/\phi_0 = 0, 1, 5,$ and 10); we also examined the cases where $\Phi_z/\phi_0$ is a half-integer, but found no clear difference. We see that various magnetic interferences appear depending on the inhomogeneity of $J_c$. The $I_c$ plot for $(x_0, y_0) = (0.2, 0.4)w$ (not shown in Fig. 5) is the same as Fig. 5(c) for $(x_0, y_0) = (0.4, 0.2)w$ with $\Phi_x$ and $\Phi_y$ interchanged.

4. **Discussion**

Figure 6 depicts the current density $J_z(x, y) = J_c(x, y)\sin[\theta(x, y)]$ across the junction plane when $I_z = I_c$ under a perpendicular magnetic field of $\Phi_z/\phi_0 = 4$, obtained from Eqs. (2), (4), and (8). In our piecewise constant $J_c$ model, as in Sec. 3.1, we have $J_z(x, y)/J_{c0} = \cos[\theta_1(x, y) - \arg(F)]$ for $x_0 < |x| < w/2$ or $y_0 < |y| < w/2$, whereas $J_z(x, y) = 0$ for $|x| < x_0$ and $|y| < y_0$. Figure 6(a) shows $J_z$ for homogeneous $J_c$ [$(x_0, y_0) = (0,0)$], and Fig. 6(b) shows $J_z$ for inhomogeneous $J_c$ [$(x_0, y_0) = (0.4, 0.4)w$].

As shown in Fig. 6(a), $J_z$ is strongly modulated near the four corners of the junction plane, but less modulated near the center of the junction plane. Consequently, the positive and negative $J_z$ do not cancel out, resulting in no magnetic interference under perpendicular magnetic fields for homogeneous $J_c$. On the other hand, as shown in Fig. 6(b), $J_z$ is modulated over the entire region of the junction plane where $J_c = J_{c0} > 0$, and positive and negative $J_z$ tend to cancel out, resulting in magnetic interference.

Contrary to our model, if $J_c$ is distributed only near the center of the junction plane, the magnetic interference is equivalent to the case where the junction plane size is small with homogeneous $J_c$. Therefore, in this case, magnetic interference does not appear under perpendicular magnetic fields.

In an overlap-type junction, as shown in Fig. 7, superconducting strips are arranged parallel to each other along the longitudinal direction (as opposed to the transverse arrangement in cross-type junctions), overlapping extremities of the strips.[4,7,8,10,11] Further details on the distinction between cross-type and overlap-type junctions, particularly regarding the magnetic interference in perpendicular magnetic fields, can be found in Refs. 10 and 11. Although we also considered the impact of inhomogeneous $J_c$ in overlap-type junctions (not included in the present paper), our examination revealed that variations in $J_c$ do not yield significant difference in the magnetic interference of overlap-type junctions.

## 5. Conclusion

We theoretically investigated the magnetic interferences of DC critical currents of cross-type junctions with inhomogeneous critical current density $J_c$ under oblique magnetic fields. We found that variations in the $J_c$ distribution generate a variety of complex magnetic interferences under oblique magnetic fields. When the $J_c$ near the edges was larger than that near the center of the junction plane, a clear magnetic interference appeared even under perpendicular magnetic fields. The mechanism of the magnetic interference was discussed. It is important that the current density near the center of the junction plane is hardly modulated under perpendicular magnetic field.

## Acknowledgment

This study has been supported by JSPS KAKENHI, Grant No. JP20K05314.

# Figure Captions

**Fig. 1.** Schematic of a cross-type junction with a barrier layer at $|x| \leq w/2$, $|y| \leq w/2$, and $|z| < d_j/2$. The top superconducting strip is located at $|x| < w/2$, $|y| < \infty$, and $d_j/2 < z < d_j/2 + d_s$. The bottom strip is located at $|x| < \infty$, $|y| < w/2$, and $-d_j/2 - d_s < z < -d_j/2$.

**Fig. 2.** Models of the $J_c$ distribution in the junction plane: $J_c = J_{c0}$ (where $J_{c0}$ is a constant) for $x_0 < |x| < w/2$ or $y_0 < |y| < w/2$, whereas $J_c = 0$ for $|x| < x_0$ and $|y| < y_0$ in the junction area. (a) Homogeneous $J_c$ with $(x_0, y_0) = (0,0)$, (b) inhomogeneous $J_c$ with $(x_0, y_0) = (0.4, 0.4)w$, (c) $(x_0, y_0) = (0.4, 0.2)w$, and (d) $(x_0, y_0) = (0.2, 0.4)w$.

**Fig. 3.** Contour plots of the critical currents $I_c(\Phi_x, 0, \Phi_z)/I_{c0}$ under 2D magnetic fields for (a) $(x_0, y_0) = (0,0)w$, (b) $(x_0, y_0) = (0.4, 0.4)w$, (c) $(x_0, y_0) = (0.4, 0.2)w$, and (d) $(x_0, y_0) = (0.2, 0.4)w$. The horizontal axis $\Phi_x/\phi_0$ is the normalized parallel magnetic field, and the vertical axis $\Phi_z/\phi_0$ is the normalized perpendicular magnetic field.

**Fig. 4.** DC critical currents under a 1D magnetic field $\Phi/\phi_0$ where $\Phi = \Phi_x$ or $\Phi = \Phi_z$ for (a) $(x_0, y_0) = (0,0)$, and (b) $(x_0, y_0) = (0.4, 0.4)w$, where the blue line represents $I_c(\Phi_x, 0, 0)/I_{c0}$, and the orange line represents $I_c(0, 0, \Phi_z)/I_{c0}$.

**Fig. 5.** Contour plots of the critical currents $I_c(\Phi_x, \Phi_y, \Phi_z)/I_{c0}$ under 3D magnetic fields for (a) $(x_0, y_0) = (0,0)w$ (top panels), (b) $(x_0, y_0) = (0.4, 0.4)w$ (middle panels), and (c) $(x_0, y_0) = (0.4, 0.2)w$ (bottom panels). The horizontal axis is $\Phi_x/\phi_0$, and the vertical axis is $\Phi_y/\phi_0$.

**Fig. 6.** Current density $J_z(x, y)$ across the junction plane for $I_z = I_c$ under a perpendicular field of $\Phi_z/\phi_0 = 4$ for (a) $(x_0, y_0) = (0,0)w$ and (b) $(x_0, y_0) = (0.4, 0.4)w$.

**Fig. 7.** Schematic of an overlap-type junction with a barrier layer sandwiched by two superconducting strips.

Figures

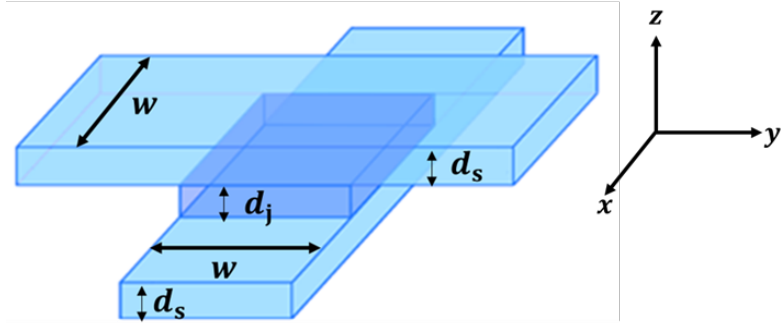

Fig. 1

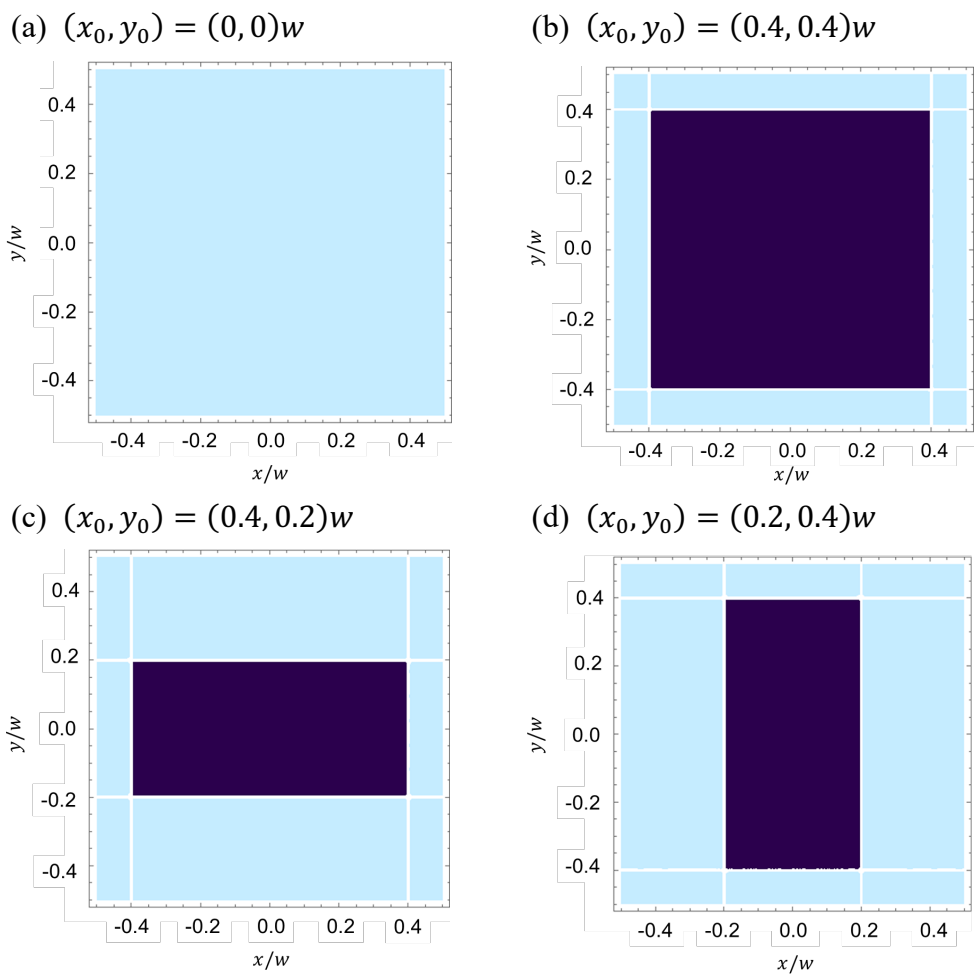

Fig. 2

(a) $(x_0, y_0) = (0, 0)w$
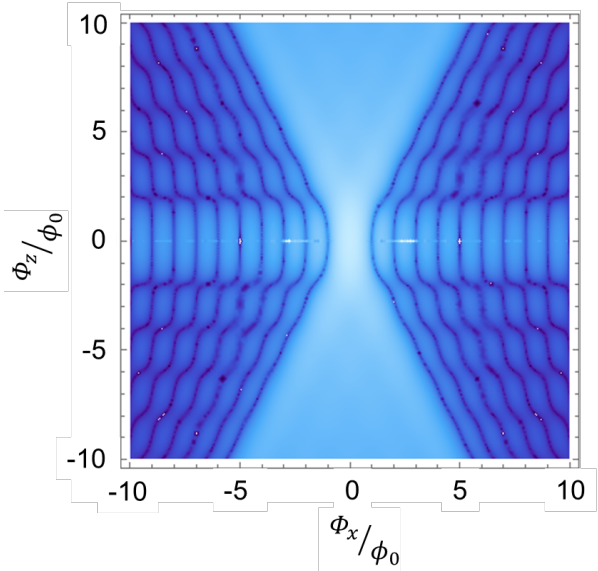

(b) $(x_0, y_0) = (0.4, 0.4)w$
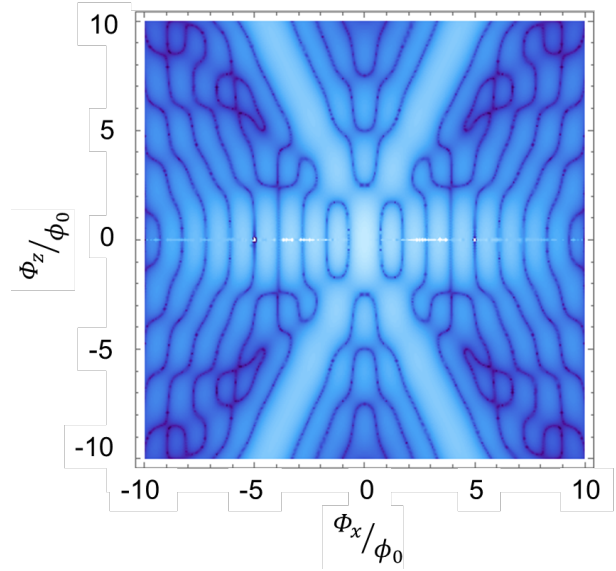

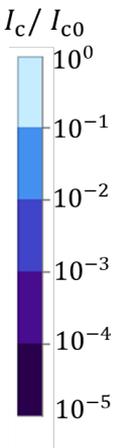

(c) $(x_0, y_0) = (0.4, 0.2)w$
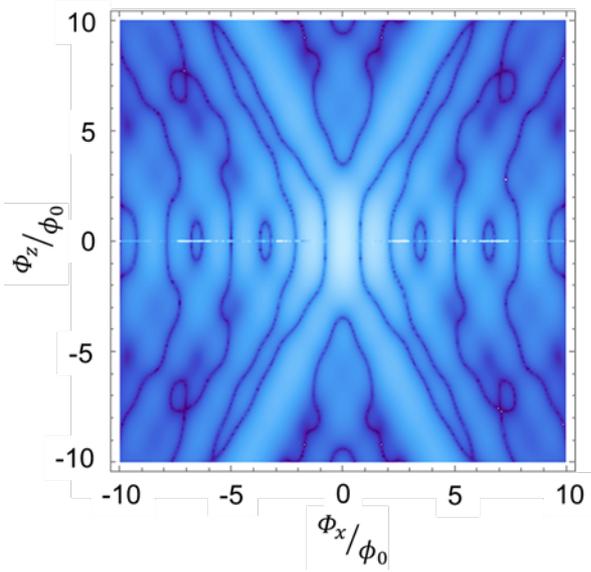

(d) $(x_0, y_0) = (0.2, 0.4)w$
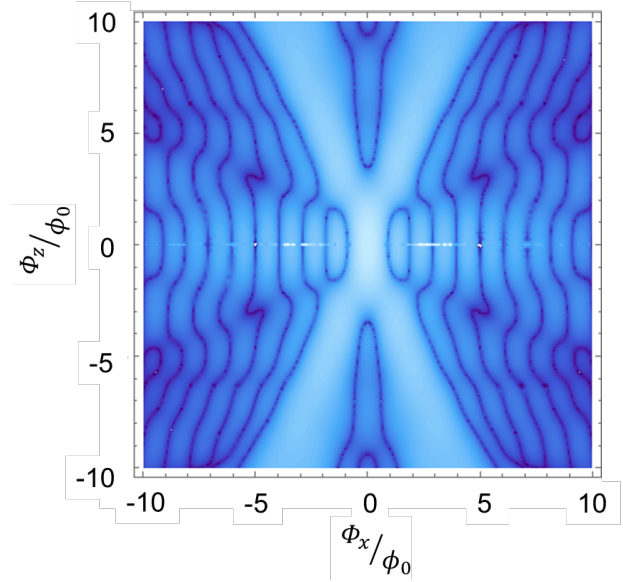

Fig. 3

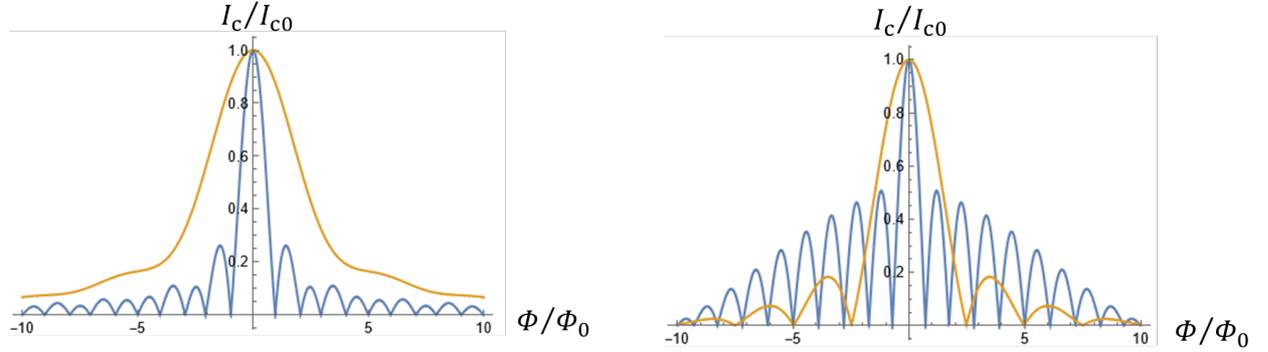

Fig. 4

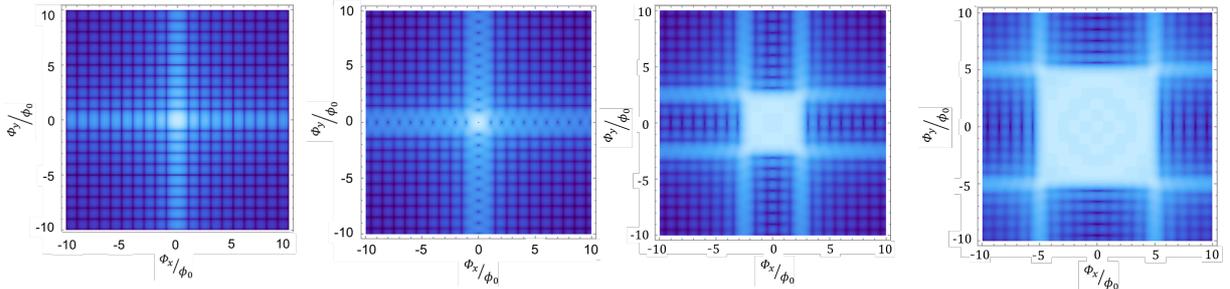

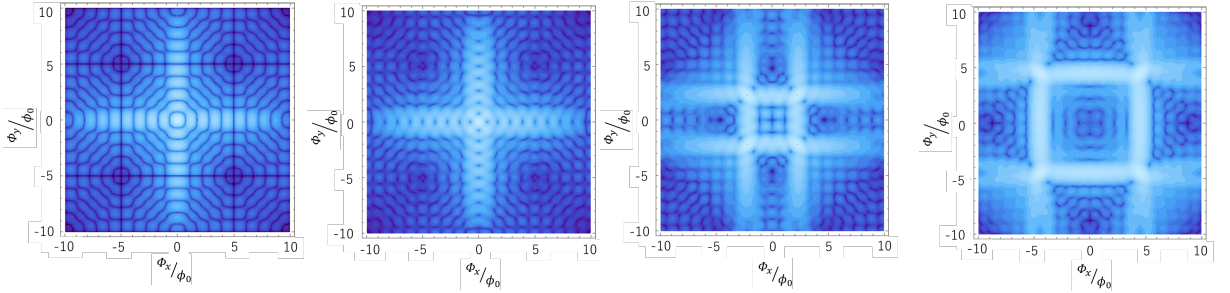

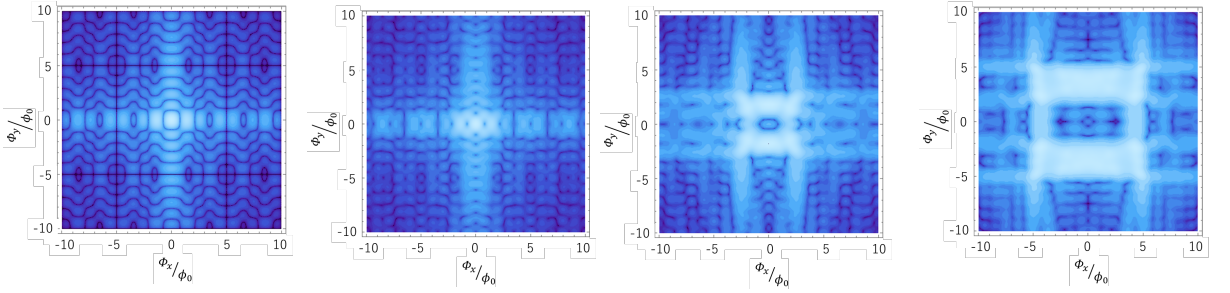

Fig. 5

(a) $(x_0, y_0) = (0, 0)w$  (b) $(x_0, y_0) = (0.4, 0.4)w$  $J_z/J_{c0}$

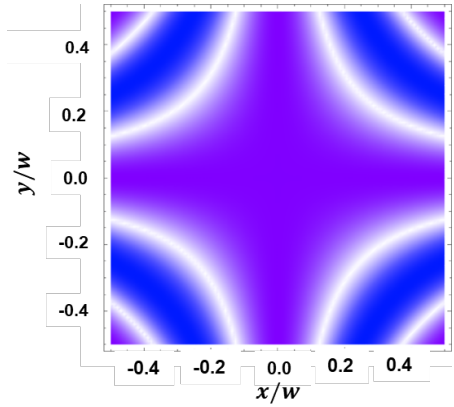
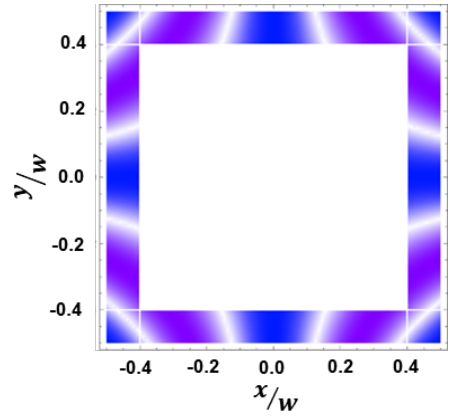
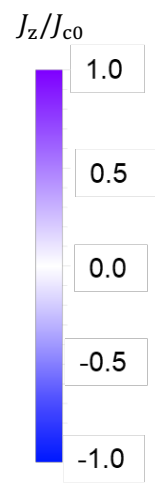

Fig. 6

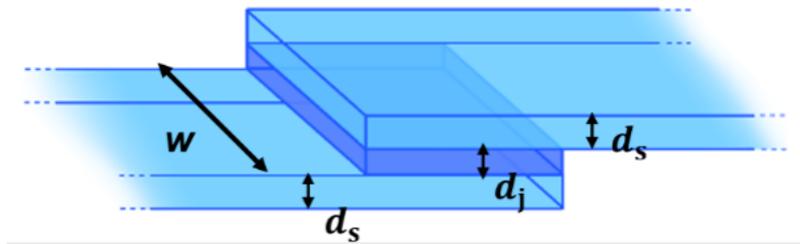

Fig. 7